\documentclass{aa}
\usepackage{epsfig}
\usepackage{amssymb}
\usepackage{times}
\usepackage{astron1}
   \thesaurus{08         % A&A Section 8: Stars,
              (08.16.4;  % Stars: AGB and post-AGB,
               08.05.3;  % Stars: evolution,
               08.09.3;)}% Stars: interiors.
   \title{On the validity of the core-mass luminosity relation
          for TP-AGB stars with efficient dredge-up }

   \titlerunning{On the validity of the core-mass luminosity relation}
 
   \author{F.\ Herwig \inst{1,2}, 
          D.\ Sch\"onberner \inst{2}
          \and 
          T.\ Bl\"ocker \inst{3}
          }
   \authorrunning{Herwig et al.\,}

   \offprints{F.\ Herwig}

   \institute{
              Universit\"at Potsdam, Institut f\"ur Physik,
              Astrophysik, Am Neuen Palais 10, 
              D-14469 Potsdam\\
              email: fherwig@astro.physik.uni-potsdam.de
         \and
              Astrophysikalisches Institut Potsdam (AIP),
              An der Sternwarte 16, D-14482 Potsdam\\
              email: fherwig@aip.de, deschoenberner@aip.de
         \and
             Max-Planck-Institut f\"ur Radioastronomie, 
             Auf dem H\"ugel 69, D-53121 Bonn\\ 
             email: bloecker@speckle.mpifr-bonn.mpg.de
             }

   \date{Received ; accepted }
\begin{document}
\maketitle

   \begin{abstract}
     We investigate the validity of the core mass - luminosity relation
     (CMLR), 
	originally described by Paczy\'nski 	
     \cite*{paczynski:70}, for
	asymptotic giant branch stars under the presence of third dredge-up
	events.
     We find, that models with efficient third dredge-up 
     with less massive cores than those associated with hot bottom
     burning \cite{bloecker:91} do not obey the linear 
     CMLR.

     Complete evolutionary calculations of thermal pulse stellar
     models which consider overshoot according to an exponential
     diffusive algorithm show systematically larger third dredge-up for
     lower core masses 
     ($0.55 \mathrm{M}_\odot \la M_\mathrm{H} \la 0.8\mathrm{M}_\odot$) 
     than any
     other existing models. We present and discuss the 
     luminosity evolution of these models. 
     
   \end{abstract}

%
%________________________________________________________________

\section{Introduction}
A general relation between the core mass and the interpulse
luminosity of asymptotic giant branch
(AGB) stars, found by Paczy\'nski \cite*{paczynski:70} as a result of
numerical computations of stellar models, 
has been widely used for the interpretation of 
observational data.
The general concept of a single linear core mass - luminosity relation (CMLR) for all AGB stars within
a certain mass range has been confirmed by succeeding evolutionary
calculations
\cite{iben:77,schoenberner:79,prwood:81,lattanzio:86,boothroyd:88b,vassiliades:93}.
Bl\"ocker and Sch\"onberner \cite*{bloecker:91}, however, 
showed that massive AGB stars 
which suffer from envelope burning do not obey this relation.
Models with a radiative zone above the H-burning shell, like those with 
lower core masses, do not show envelope burning 
and can be described by relations like 
\cite{bloecker:93}: 
\begin{equation}
\label{glei:cmlr-bloecker93}
L/\mathrm{L}_\odot=62200(M_\mathrm{H} / \mathrm{M}_\odot - 0.487) \mathrm{ \hspace*{0.1cm},}
\end{equation}
where 
$0.55 \mathrm{M}_\odot \la M_\mathrm{H} \la 0.8\mathrm{M}_\odot$
is the  mass of the hydrogen-exhausted
core and $L$ is the maximum  pre-flash luminosity.
Such linear relations have been derived from
models which do not show any considerable third dredge-up.
Note, that efficient dredge-up is required in order to explain the observational
luminosity function of carbon stars (e.g.\ Marigo \mbox{et al.}\,\cite*{marigo:96}).
 
It has been reported by Herwig et al.\,\cite*{herwig:97} that the
application of exponential diffusive overshoot, which is
motivated by results of hydrodynamical simulations of convectively
unstable surface layers \cite{freytag:96}, leads to stronger third
dredge-up events than previously found. We have further
investigated the impact of this kind of overshoot on AGB stellar
models.  A detailed account of the new, more extended calculations
will be given elsewhere.  In this {\it Letter} the
luminosity evolution of models with lower core masses and
efficient dredge-up events is described.
We show that strong dredge-up leads to the violation of the CMLR
for models that have lower core masses than those 
associated with envelope burning.
We compare our results with predictions obtained by the homology relations.

%__________________________________________________________________
\section{The models}
\label{sec:the_models}
\begin{figure*}
  \hbox{\hspace{0cm}\epsfxsize=\textwidth
    \epsfbox{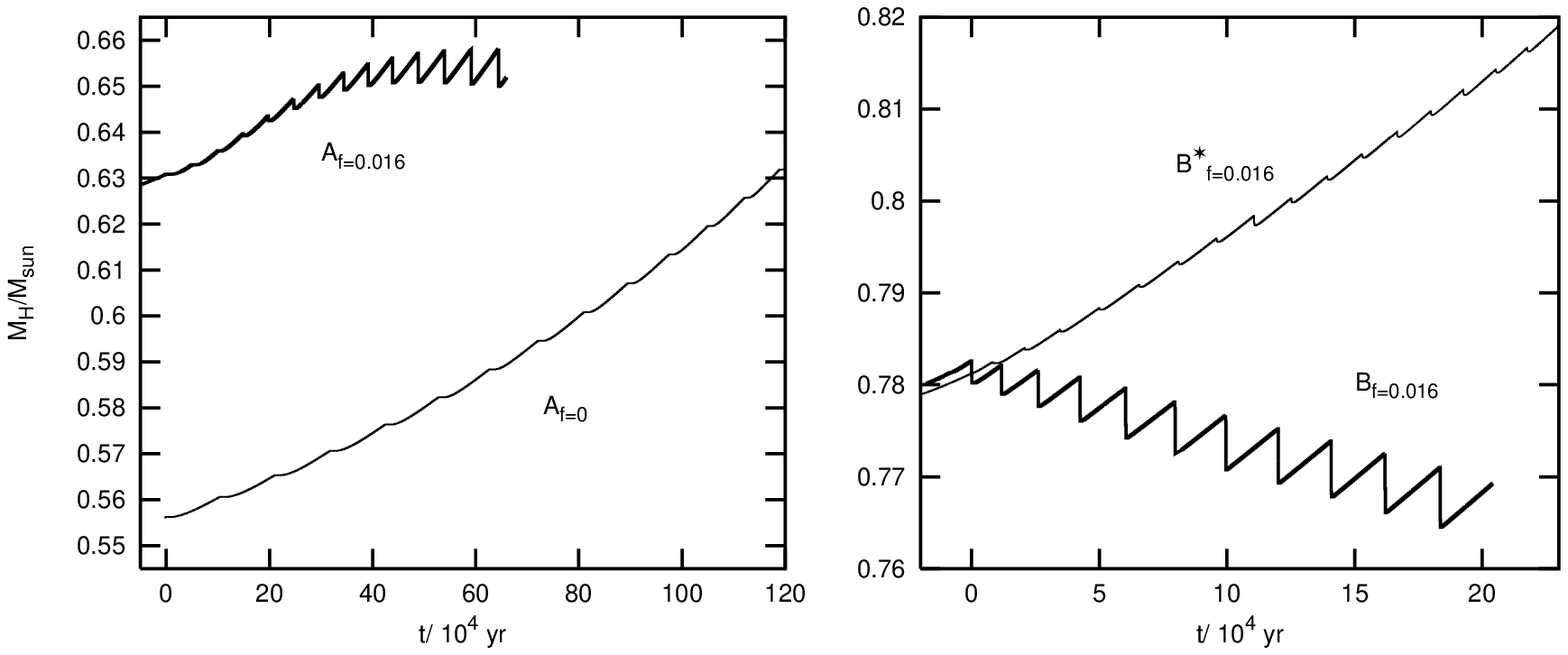}}
\caption[]{\label{M_H-paper4} Evolution of the core mass during the thermal
	pulse - AGB
  evolution of model sequences with initial masses of $3\mathrm{M}_\odot$ 
	(left, label \textsf{A}) and
  $4\mathrm{M}_\odot$ (right, label \textsf{B}). 
        The subscript indicates the treatment of overshoot: 
	$\ensuremath{\mathrm{f=0}}$ - no overshoot, $\ensuremath{\mathrm{f=0.016}}$  overshoot with the
	given efficiency
	according to the method described in the text. The $^\bigstar$ denotes
	a sequence which has been computed with reduced
	numerical resolution. While overall TP-AGB evolution is
	not strongly affected by this, the third dredge-up 
	is weaker by an order of magnitude (see text).}
\end{figure*}
The evolutionary code and the physical parameters 
[(Y,\,Z,\,$\alpha_\ensuremath{\mathrm{MLT}}$)=(0.28,\,0.02,\,1.7)] are the same as the
one used by Herwig et al.\,\cite*{herwig:97}.  Mixing has been treated in
a time dependent manner by solving a diffusion equation for each
isotope.  The diffusion coefficient $D$ depends on the assumed mixing
model.
For the regions which are
immediately adjacent to convectively unstable zones we allow for
overshooting by using a depth dependent diffusion coefficient. It
has been derived by Freytag et al.\,(1996: Eq.\,9) \nocite{freytag:96}
from their numerical simulations of two-dimensional radiation
hydrodynamics of time-dependent compressible convection.  In this
overshoot prescription the coefficient $f$ is a measure of the
efficiency of the exponentially declining diffusive mixing beyond the
boundary of convection.  The findings presented in this paper are
based on AGB models which have been computed with $f=0.016$.
If applied to core convection, this value of $f$ 
reproduces the observed width of the
main sequence \cite{schaller:92}. A detailed discussion is given by
Herwig \cite*{herwig:98a}. Standard models, which have been
computed for comparison, are referred to as $f=0$ models.

Before we comment on the core mass evolution of our model sequences
(Fig.\,\ref{M_H-paper4}), recall, that two main processes are altering the
core mass: (a) hydrogen shell burning during the interpulse phase
increases the mass while (b) the third dredge-up occurring
shortly after the thermal pulse (TP) reduces the mass.

We have computed two AGB model sequences with $f=0.016$: at the first
TP sequence \ensuremath{\mathrm{A}_\mathrm{f=0.016}}
 exhibits a core mass $\ensuremath{M_{\rm H1}}=0.631\mathrm{M}_\odot$ and sequence \ensuremath{\mathrm{B}_\mathrm{f=0.016}} has
$\ensuremath{M_{\rm H1}}=0.783\mathrm{M}_\odot$.  
Sequence \ensuremath{\mathrm{A}_\mathrm{f=0.016}} experiences dredge-up from the third TP on with steadily
increasing efficiency (Fig.\,\ref{M_H-paper4}). 
At the eleventh TP the dredge-up parameter
$\lambda$\footnote{$\lambda$ is defined as the ratio of  
core mass decrease by
dredge-up to core mass growth by hydrogen burning during the 
TP cycle.} 
has reached unity. Already at the first pulse of sequence \ensuremath{\mathrm{B}_\mathrm{f=0.016}} 
the dredge-up parameter exceeds unity. Starting from $\lambda=1.6$ it
gradually decreases to $1.3$ at the last TP computed so far.  This
enhanced efficiency of dredge-up, compared to existing
calculations, is caused by the application of the exponential
diffusive overshoot algorithm and will be further described in a
forthcoming paper.  The  $f=0$ sequence
\ensuremath{\mathrm{A}_\mathrm{f=0}} with $\ensuremath{M_{\rm H1}}=0.556\mathrm{M}_\odot$ shows no
third dredge-up (Fig.\,\ref{M_H-paper4}).  
The sequence \ensuremath{\mathrm{B}_\mathrm{f=0.016}^\bigstar} has been computed with
identical assumptions as sequence \ensuremath{\mathrm{B}_\mathrm{f=0.016}} with the
only difference that our improved adaptive step and grid size
algorithm \cite{herwig:98a} has not been applied. As a result, the
dredge-up remains weak ($\lambda \simeq 0.12$).  Sequence \ensuremath{\mathrm{B}_\mathrm{f=0.016}^\bigstar} serves
only for comparison with sequence \ensuremath{\mathrm{B}_\mathrm{f=0.016}} as an example of a TP-AGB
evolution with only weak dredge-up but otherwise identical properties.

None of the models presented here shows envelope burning 
(hot bottom burning, HBB) because the core masses are 
too low.

%__________________________________________________________________
\section{The core mass - luminosity relation}
\label{sec:the_cmlr}

\subsection{Results from the models}

\begin{figure}
\epsfxsize=8.8cm \epsfbox{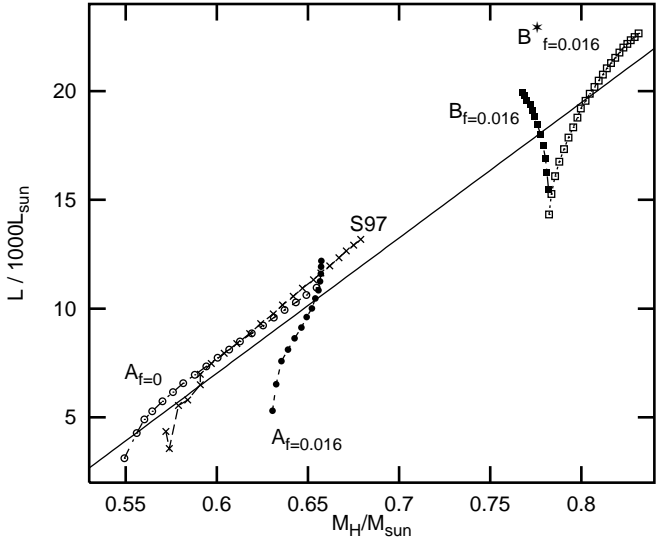}
\caption[]{\label{M_H-L-paper4} Interpulse luminosity vs.\ core mass
  for the two A and B sequences introduced in Sec.\,\ref{sec:the_models}.
  Each symbol represents one TP. \textsf{S97} (crosses)
  refers to a $3\mathrm{M}_\odot$ AGB model
  sequence of Straniero et al.\,(1997). The solid line represents
  Eq.\,\ref{glei:cmlr-bloecker93}.}
\end{figure}

In Fig.\,\ref{M_H-L-paper4} we show the core masses and  
luminosities for each computed TP. 
The values refer to the end of the interpulse period when the
luminosity is predominantly generated in the hydrogen burning shell
and has reached its maximum.

Due to  dredge-up the core-mass growth $\Delta \ensuremath{M_{\rm H}}$
per TP  is smaller compared to sequences
without dredge-up or even negative. 
The luminosities and core masses of the model sequences with 
efficient dredge-up (lines with filled symbols in Fig.\,\ref{M_H-L-paper4}) 
are not reproduced by the linear  CMLR 
(solid line, Eq.\,\ref{glei:cmlr-bloecker93}). They violate the classical
relation.
  
During the first few pulses of sequence  \ensuremath{\mathrm{A}_\mathrm{f=0.016}} dredge-up is negligible.
The relation bends asymptotically towards the linear relation (solid line,
Eq.\,\ref{glei:cmlr-bloecker93}).  As the dredge-up gains strength from pulse
to pulse this trend turns into an upwards inflection. At the last
pulses, shown in Fig.\,\ref{M_H-L-paper4}, the dredge-up parameter slightly 
exceeds unity (\ensuremath{M_{\rm H}} does not
grow anymore) and the luminosity is still increasing.  For sequence
\ensuremath{\mathrm{B}_\mathrm{f=0.016}} the dredge-up parameter always clearly exceeds unity.  
Accordingly
the core mass is \emph{reduced} after each TP.  However, the
corresponding luminosities are still \emph{increasing} (Fig.\,\ref{M_H-L-paper4}).
This CMLR runs almost perpendicular to the linear relation.
\nocite{straniero:97}

The relation of sequence \ensuremath{\mathrm{A}_\mathrm{f=0}} (no third dredge-up) runs
parallel to the relation given by Eq.\,\ref{glei:cmlr-bloecker93}. The small
offset is due to the different opacities used.
The relation labeled \textsf{S97} in
Fig.\,\ref{M_H-L-paper4} has been taken from Straniero
et al.\,cite*{straniero:97}. Their models exhibit dredge-up with
$\lambda<0.46$, leading only to a slight upturn.
We conclude that \emph{with
  the occurrence of sufficiently efficient dredge-up the AGB models
  do not follow the classical core mass -- luminosity relation
  anymore.} As will be explained below, this 
  deviation from the linear CMLR has a different physical
  origin than the one due to HBB for massive 
  AGB stars.

\subsection{Explanation from homology relations}

Kippenhahn \cite*{kippenhahn:81}, extending previous works by Refsdal
\& Weigert \cite*{refsdal:70}, has demonstrated how homology relations
can provide an expression for the luminosity of shell burning
stellar models. E.g.\ the homology relation for the luminosity
reads 
\begin{equation}
   \label{homo-l}
   l(\frac{r}{R_\ensuremath{\mathrm{H}}}) \approx
   \ensuremath{M_{\rm H}}^\ensuremath{\mathrm{\sigma_1}}R_\ensuremath{\mathrm{H}}^\ensuremath{\mathrm{\sigma_2}} \mathrm{ \hspace*{0.1cm}.} 
\end{equation} 
Here $l$ and
$r$ are the luminosity and radius at homologous points of two
models. $R_\ensuremath{\mathrm{H}}$ and \ensuremath{M_{\rm H}} are the core radius and the core mass. 
Thus, in the
case of an interpulse AGB model shortly before the next TP,
$l$ at the upper boundary of the shell is
practically the surface luminosity.  Kippenhahn
\cite*{kippenhahn:81} has derived for the exponents of Eq.\,\ref{homo-l}
the expressions
\begin{equation}
\label{gl:sigma}  
 \sigma_1=\frac{4n+\nu}{N} \mathrm{ \hspace*{0.1cm},} \ensuremath{\mathrm{\ }}  \sigma_2=\frac{3-\nu-3n}{N}\beta 
\end{equation} 
with 
\begin{equation}
\label{gl-N}  
 N=(4-3\beta)(1+n)+(1-\beta)(\nu-4) \mathrm{ \hspace*{0.1cm},}
\end{equation} 
where $\beta:=\frac{P_\ensuremath{\mathrm{gas}}}{P}$ is the gas pressure fraction
and $a$, $b$, $n$, and $\nu$ are the exponents of the power-law
approximations for the opacity and the energy production rate: 
\begin{equation}
 \kappa=\kappa_\ensuremath{\mathrm{0}}P^aT^b \mathrm{ \hspace*{0.1cm},} \ensuremath{\mathrm{\ }}\epsilon =
 \epsilon_\ensuremath{\mathrm{0}}\rho^{n-1}T^\nu  \mathrm{ \hspace*{0.1cm}.}
\end{equation}

Suitable estimates for the exponents in the
approximations for $\kappa$ and $\epsilon$ are $a=b=0$
 for electron
scattering and $\nu=14$, $n=2$ for CNO cycle. For
$\beta \rightarrow0$ 
(radiation pressure is dominant, large core masses) the dependence
of the luminosity on the core radius decreases ($\sigma_2=0$) and
the homology method predicts a linear CMLR ($\sigma_1=1$).  However,
it has been shown by Bl\"ocker and Sch\"onberner \cite*{bloecker:91}
that models with core masses larger than $\ensuremath{M_{\rm H}} \approx 0.8\mathrm{M}_\odot$ 
violate the
linear CMLR anyway due to a different reason: the convective
envelope region can penetrate into the hydrogen burning shell
which leads to enhanced H-burning.

In the range of core masses considered here ($0.55 \la \ensuremath{M_{\rm H}}/\mathrm{M}_\odot \la 0.8$)
radiation pressure is smaller ($\beta$ is larger) and the dependence
of $L$ on $R_\ensuremath{\mathrm{H}}$ remains important (while HBB does not occur).
For $\beta=0.6$ (which roughly corresponds to $\ensuremath{M_{\rm H}} \approx 0.8\mathrm{M}_\odot$)
the homology method predicts the core mass - core radius - luminosity
relation
\begin{equation}
\label{L-M-middle}
 L \propto \ensuremath{M_{\rm H}}^2R_\ensuremath{\mathrm{H}}^{-1} \mathrm{ \hspace*{0.1cm}.} 
\end{equation}
\begin{figure}
  \epsfxsize=8.8cm \epsfbox{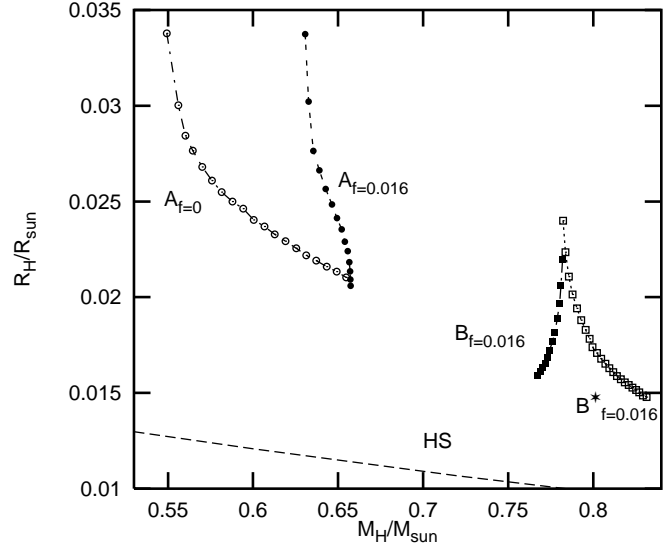}
\caption[]{\label{Mc-Rc-paper4} 
  Core radius ($R_\ensuremath{\mathrm{H}}$) vs.\ core mass for the sequences described
  in Sec.\,\ref{sec:the_models}. The radius is largest at the first TP.  The
  line \textsf{HS} gives the mass -- radius relation of Hamada and Salpeter
  (1961) for white dwarfs.  }
\end{figure}
\nocite{hamada:61}
Thus, for models of lower core mass the homology
method cannot predict any CMLR without an additional
relation between core mass and core radius. 
I.\,e.\ \emph{the luminosity is not a function
of the core mass alone.}

Therefore, we display the core radii vs.\ core masses for 
our model sequences in Fig.\,\ref{Mc-Rc-paper4}. 
Sequence \ensuremath{\mathrm{A}_\mathrm{f=0}} shows a strong core radius
decrease over the first few pulses when the core mass 
growth is rather small.
Afterwards, the core mass - core radius relation (CMCRR) is  almost linear.
The models of sequences \ensuremath{\mathrm{A}_\mathrm{f=0.016}} and \ensuremath{\mathrm{B}_\mathrm{f=0.016}} 
also show a steady radius decrease, although their 
core masses evolve quite differently.
For example, when the dredge-up 
parameter $\lambda$ has reached unity  
($\Delta \ensuremath{M_{\rm H}} /\mathrm{M}_\odot =0$, sequence \ensuremath{\mathrm{A}_\mathrm{f=0.016}}) the
core radius continues to decrease steadily from pulse to pulse 
leading to a 
vertical CMCRR.  
This leads, according to Eq.\,\ref{L-M-middle}, to continuously increasing 
luminosities at constant core mass which in turn leads to the 
deviation from the linear CMLR.
Also for sequence \ensuremath{\mathrm{B}_\mathrm{f=0.016}} and \ensuremath{\mathrm{B}_\mathrm{f=0.016}^\bigstar}, Fig.\,\ref{Mc-Rc-paper4} shows a steady core radius
decrease per pulse  despite the large difference in both $\lambda$ and 
core mass evolution.  The CMCRR of a model sequence
with efficient dredge-up is most importantly determined by the
dependence of the core mass evolution on the dredge-up parameter.

From Eq.\,\ref{L-M-middle} the importance of
the CMCRR for the understanding of the deviations from
the linear CMLR found in the models with efficient dredge-up
is evident.
It is illustrative to insert the core
masses and core radii of the sequences \ensuremath{\mathrm{B}_\mathrm{f=0.016}} and \ensuremath{\mathrm{B}_\mathrm{f=0.016}^\bigstar} 
(see Fig.\,\ref{Mc-Rc-paper4}) into Eq.\,\ref{L-M-middle}.
\begin{figure}
  \epsfxsize=8.8cm 
  \epsfbox{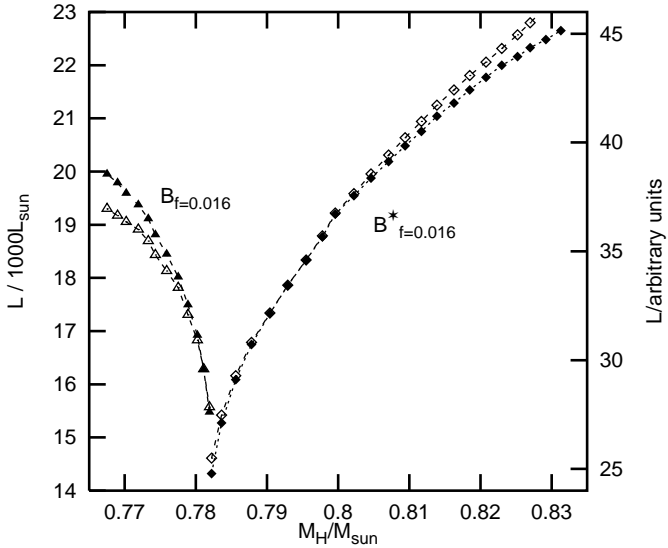}
\caption[]{\label{HOMO-4M-paper4} 
  Interpulse luminosities of the full numerical model sequences \ensuremath{\mathrm{B}_\mathrm{f=0.016}} 
  and \ensuremath{\mathrm{B}_\mathrm{f=0.016}^\bigstar} in comparison with the ''semi-analytical'' luminosities.
  The filled symbols repeat the respective lines from
  Fig.\,\ref{M_H-L-paper4} and refer to the left ordinate. The open symbols
  show the values according to Eq.\,\ref{L-M-middle} if the core radius and
  core mass of the numerical models (as shown in Fig.\,\ref{Mc-Rc-paper4})
  are inserted.}
\end{figure}
In Fig.\,\ref{HOMO-4M-paper4} one can see that the resulting
``semi-analytical'' luminosities do reproduce the functional
dependence of luminosity and core mass for both cases.  
The fact that the homology relations reproduce the deviations
from the CMLR illustrates also the different physical origin
of this deviation compared to the previously mentioned one
due to HBB \cite{bloecker:91}.

%__________________________________________________________________
\section{Conclusions}
\label{sec:conclusions}
We have shown that models with efficient dredge-up and core masses
smaller than associated with HBB, do not obey the Paczy\'nski 
core mass -- luminosity relation because their core
evolution with respect to its mass and radius is so
different.

Our finding does not depend on the precise physical
circumstances under which the third dredge-up occurs.
The deviations 
from the CMLR reported here should be a general feature
applying to all stars which are believed to suffer efficient
dredge-up (and no HBB).
Thus, it should be critically
reviewed that synthetic models assume both considerably efficient dredge-up 
\emph{and} a linear core mass -- luminosity relation
to be valid \cite{marigo:96,groenewegen:95}.
Another immediate consequence is that core masses based on 
observed luminosities and the CMLR are
probably too large. Finally, the deviation from the CMLR described 
here should be preserved during the post-AGB evolution, in contrast
to that due to HBB. A
 physically related situation of post-AGB stars with identical
core mass and different initial masses has been studied by
 Bl\"ocker and Sch\"onberner \cite*{bloecker:90} and Bl\"ocker 
\cite*{bloecker:95b}. The different
history leads to remnants with different core radii and consequently 
different luminosities.

Since the pioneering calculations of Paczy\'nski 	
     \cite*{paczynski:70} the 
core-mass luminosity  relation has been widely used and became
a basic ingredient for many applications.
However, it seems to turn out that the existence of such a general relation
should be doubted for the large fraction of AGB stars which suffer
either hot bottom burning or efficient dredge-up.

%__________________________________________________________________
\begin{acknowledgements}
  F.\ H.\ and T.\ B.\ acknowledge funding by the \emph{Deut\-sche
    For\-schungs\-ge\-mein\-schaft, DFG\/} (grants Scho\,394/13, Ko\,738/12
    and La\,587/16).
\end{acknowledgements}

%__________________________________________________________________

\end{document}